\documentclass[12pt]{iopart}

\usepackage{iopams}  
\usepackage{graphics}
\begin{document}

\title[Constructive proof of the Kerr-Newman black hole uniqueness]{Constructive
proof of the Kerr-Newman black hole uniqueness including
the extreme case}

\author{Reinhard Meinel}

\address{Theoretisch-Physikalisches Institut,
University of Jena,\\
 Max-Wien-Platz 1, 07743 Jena, Germany}

\ead{meinel@tpi.uni-jena.de}

\begin{abstract}
A new proof of the uniqueness of the Kerr-Newman black hole solutions amongst
asymptotically flat, stationary and axisymmetric electro-vacuum spacetimes
surrounding a connected Killing horizon is given by means of an explicit
construction of the corresponding complex Ernst potentials on the axis of
symmetry. This construction, which makes use of the inverse scattering method,
also works in the case of a degenerate horizon.  
\end{abstract}


\vspace{2pc}

\section{Introduction}
Recently, several attempts have been made to extend the well-known uniqueness
results for the Kerr \cite{carter71, robinson75, cc08} and Kerr-Newman black
holes \cite{mazur82, bunting83, costa10} to the previously excluded case of a
degenerate horizon leading to the {\it extremal} Kerr or Kerr-Newman black
holes. In \cite{rfe} (end of section 2.4) it was shown that the constructive
Kerr uniqueness proof given in \cite{neugebauer00, nm03}, which assumes
stationarity {\it and} axial symmetry from the very beginning and makes use of
the ``inverse scattering method'', see also the earlier work \cite{varzugin97},
can indeed be extended to the degenerate case. In 2010 further proofs of the
extremal Kerr \cite{fl10} and extremal Kerr-Newman uniqueness \cite{ahmr10,
cn10} were published. It is beyond the scope of the present paper to analyse
the various mathematical assumptions of these proofs. Instead, a generalization
of the above mentioned constructive Kerr uniqueness proof
\cite{neugebauer00, nm03} to the Kerr-Newman case, again including the
possibility of degenerate horizons as in \cite{rfe}, is presented. The paper is 
organized as follows: In Section 2, we briefly describe the Ernst 
formulation \cite{ernst68} of the relevant field
equations and the ``Linear Problem'' introduced by Neugebauer and Kramer
\cite{nk83}, which is used for the proof. Section 3 deals with the integration
of the Linear Problem along the axis of symmetry and along the horizon. Here the
horizon boundary conditions defining a stationary black hole, see \cite{he73,
carter73}, enter into the procedure. In Section 4, the uniqueness proof is
completed by closing the integration path via infinity and deriving explicit
expressions for the complex Ernst potentials on the axis of symmetry.
Some general remarks on solutions showing reflection symmetry with
respect to an ``equatorial plane'' can be found in Section 5, and in Section
6 we discuss further possible applications of the method. 

\section{Field equations and Linear Problem}
The Einstein-Maxwell vacuum equations (governing ``electro-vacuum spacetimes''),
in the case of stationarity and axisymmetry, can be reduced to a system of
coupled non-linear partial differential equations for the complex Ernst
potentials $\mathcal E(\rho,\zeta)$ and $\Phi(\rho,\zeta)$ \cite{ernst68}: 
\begin{equation}\label{ernst}
f\,\nabla^2 {\mathcal E}=(\nabla {\mathcal E} +
2\bar{\Phi}\nabla \Phi)\cdot\nabla {\mathcal E}\, , \quad
f\,\nabla^2 \Phi=(\nabla {\mathcal E} +2\bar{\Phi}\nabla \Phi)\cdot\nabla \Phi
\end{equation}
with
\begin{equation}\label{f}
f\equiv\Re\, {\mathcal E} + |\Phi|^2\, ,
\end{equation}
where the operator $\nabla$ has the same meaning as in Euclidean 3-space in
which $\rho$, $\zeta$ and $\phi$ are cylindrical coordinates. The spacetime line
element is 
\begin{equation}\label{line}
{\rm d}s^2=f^{-1}[\,h({\rm d}\rho^2+{\rm d}\zeta^2)+\rho^2{\rm d}\phi^2]
-f({\rm d}t+a\,{\rm d}\phi)^2
\end{equation}
or, equivalently,
\begin{equation}\label{line2}
{\rm d}s^2=\e^{2\alpha}({\rm d}\rho^2+{\rm d}\zeta^2)+\rho^2\e^{-2\nu}
({\rm d}\phi - \omega\,{\rm d}t)^2-e^{2\nu}{\rm d}t^2\, ,
\end{equation}
the latter form beeing of advantage in the presence of ergospheres. 
Here $f(\rho,\zeta)$ is directly given by (\ref{f}). The other metric functions
$h(\rho,\zeta)$ and $a(\rho,\zeta)$ as well as the electromagnetic field can be
calculated in a straightforward way from $\mathcal E$ and $\Phi$, see
\cite{ernst68, es}. The relations between $f$, $h$, $a$ and $\alpha$, $\nu$, 
$\omega$ can be read off by comparing (\ref{line}) and (\ref{line2}). 
The coordinates $t$ (time) and $\phi$ (azimuthal angle) are adapted to the 
Killing vectors
\begin{equation}\label{killing}
\boldsymbol\xi=\frac{\partial}{\partial t}\, , 
\quad \boldsymbol\eta=\frac{\partial}{\partial \phi}
\end{equation}
corresponding to stationarity and axial symmetry.
We assume asymptotic flatness, i.e.\ $f\to 1$, $h\to 1$ and $a\to 0$ at spatial 
infinity. The asymptotic behaviour of $\mathcal E$ and $\Phi$ is given by
\begin{equation}\label{as1}
\Re\,{\mathcal E} = 1 - \frac{2M}{r} + {\mathcal O}(r^{-2})\, , 
\quad \Im\, {\mathcal E} = -\frac{2J\cos\theta}{r^2} + {\mathcal O}(r^{-3})\, , 
\end{equation}
\begin{equation}\label{phi}\label{as2}
\quad \Phi = \frac{Q}{r} + {\mathcal O}(r^{-2}) 
\end{equation}
with
\begin{equation}\label{sc}
\rho=r\sin\theta\, , \quad \zeta=r\cos\theta\, .
\end{equation}  
The multipole moments $M$, $J$ and $Q$ are the gravitational mass, the 
($\zeta$-component of the) angular momentum and the electric charge. Note that 
we exclude magnetic monopoles here.\footnote{A magnetic monopole can be
reintroduced 
in the end by means of a duality rotation.}

Equations (\ref{ernst}) turned out to be integrable in the sense of soliton
theory, i.e., there exists a system of linear differential equations (called
Linear Problem) having  (\ref{ernst}) as its integrability condition,
cf.~\cite{bv01}. We use here a slightly modified version of the Linear Problem
(LP) presented by Neugebauer and Kramer \cite{nk83}: 
\begin{equation}\label{LP1}
{\bf Y}_{,z}=\left[\left(\begin{array}{ccc}
B_1 & 0 & C_1 \\
0 & A_1 & 0 \\
D_1 & 0 & 0
\end{array}
\right)+\lambda\left(\begin{array}{ccc}
0 & B_1 & 0 \\
A_1 & 0 & -C_1 \\
0 & D_1 & 0
\end{array}
\right)\right]{\bf Y}\, ,
\end{equation}
\begin{equation}\label{LP2}
{\bf Y}_{,\bar{z}}=\left[\left(\begin{array}{ccc}
B_2 & 0 & C_2 \\
0 & A_2 & 0 \\
D_2 & 0 & 0
\end{array}
\right)+\frac{1}{\lambda}\left(\begin{array}{ccc}
0 & B_2 & 0 \\
A_2 & 0 & -C_2 \\
0 & D_2 & 0
\end{array}
\right)\right]{\bf Y}
\end{equation}
with
\begin{equation}\label{lambda}
\lambda=\sqrt{\frac{K-{\rm i}\bar z}{K+{\rm i}z}}\, ,
\end{equation}
\begin{equation}
z=\rho+{\rm i}\zeta\, , \quad \bar z=\rho-{\rm i}\zeta\, ,
\end{equation}
\begin{equation}\label{AB}
A_1=\bar{B}_2=\frac{{\mathcal E}_{,z}+2\bar{\Phi}\Phi_{,z}}{2f}\, , \quad
A_2=\bar{B}_1=\frac{{\mathcal E}_{,\bar z}+2\bar{\Phi}\Phi_{,\bar z}}{2f}\, ,
\end{equation}
\begin{equation}\label{CD}
C_1=f\bar{D}_2=\Phi_{,z}\, , \quad C_2=f\bar{D}_1=\Phi_{,\bar z}\, ,
\end{equation}
a bar denoting complex conjugation.
Note that the $3\times 3$ matrix ${\bf Y}$ is related to the matrix 
${\bf \Omega}$ in \cite{nk83} by
the simple transformation
\begin{equation}\label{trans}
{\bf Y}=\left(\begin{array}{ccc}
1 & 0 & 0 \\
0 & 1 & 0 \\
0 & 0 & {\rm i}f^{-1/2}
\end{array}
\right){\bf \Omega}\, .
\end{equation}
The use of ${\bf Y}$ instead of ${\bf \Omega}$ avoids some problems in the
presence of ergospheres, where the function $f$ becomes negative. In
particular, the expressions (\ref{AB}, \ref{CD}) for the coefficients $A_i$,
$B_i$, $C_i$ and $D_i$ ($i=1,2$) in the LP (\ref{LP1}, \ref{LP2}), in contrast 
to the corresponding expressions in \cite{nk83}, do not contain
square roots of $f$. This ensures that relations like $A_2=\bar{B}_1$ remain unchanged
inside the ergosphere and no ambiguities occur. The
boundaries of ergospheres, where $f=0$ holds, require a separate discussion. 
At these ``ergosurfaces'' some of the coefficients in the LP are singular. The same 
holds for the differential relations that have to be used for calculating the full
metric and the electromagnetic field from $\mathcal E$ and $\Phi$. However, 
it can easily be verified from these relations that a smooth spacetime metric 
(together with a smooth electromagnetic field), 
which we assume, leads to smooth potentials $\mathcal E$ and $\Phi$ at ergosurfaces 
(for $\rho\ne 0$). In the opposite direction, the Ernst equations themselves 
provide conditions which ensure smoothness (and the Lorentzian character) of the
metric and smoothness of the electromagnetic field at ergosurfaces. Without 
electromagnetic field this was studied in detail in \cite{cgms06} and a partial 
treatment of the electro-vacuum case can be found in \cite{cs08}.
In the case of a zero of first order 
($f=0$, $\nabla f\ne 0$) smooth potentials $\mathcal E$ and $\Phi$ satisfy 
either (i) $\mathcal E_{,z}=0$, $\Phi_{,z}=0$ or (ii) $\mathcal E_{,\bar z}=0$, 
$\Phi_{,\bar z}=0$ as a consequence of the vanishing right-hand sides in (\ref{ernst}).
In both cases, regularity of the metric functions $\alpha$, $\nu$ and $\omega$ in 
(\ref{line2}), calculated (via $f$, $h$, $a$) from $\mathcal E$ and $\Phi$, is assured.
In case (i), the only singular coefficients in the LP are $A_2$, $B_1$ and $D_1$. 
Regularity of ${\bf Y}_{,z}$ and ${\bf Y}_{,\bar z}$ results from the relation 
$Y_{1k}+\lambda Y_{2k}=0$ ($Y_{ik}$ denoting the elements of the matrix ${\bf Y}$), 
which is consistent with 
${\rm det}\,{\bf Y}=0$ at the ergosurface, see equation (\ref{det}) below. 
Analogously, in case (ii), with singular coefficients $A_1$, $B_2$ and $D_2$, we have 
$\lambda Y_{1k}+Y_{2k}=0$. In our proof we are not making any a priori assumptions
on the shape and other properties of ergosurfaces. However, the resulting ergosurface 
(of the Kerr-Newman black hole) is indeed characterized by a first-order zero of $f$. 
The points where the horizon meets the axis of symmetry are discussed 
at the end of subsection \ref{ha}. 

It is important to note that ${\bf Y}$ depends not only on the coordinates
$\rho$ and $\zeta$ (or $z$ and $\bar z$), but also on the additional complex
``spectral parameter'' $K$, which enters the LP (\ref{LP1}, \ref{LP2}) via
$\lambda$ as given in (\ref{lambda}). Since 
$\bar K$ does not appear, we can assume without loss of generality that the
elements of 
${\bf Y}$ are holomorphic functions of $K$ defined on the two-sheeted Riemann
surface associated with (\ref{lambda}). For a given solution $\mathcal E$, 
$\Phi$ to the Einstein-Maxwell equations, the solution to the LP can be fixed by
prescribing ${\bf Y}$ at some point $\rho_0$, $\zeta_0$ of the $\rho$-$\zeta$
plane as a (matrix) function of $K$ in one of the two sheets of the
Riemann
surface.\footnote{The values of the elements of ${\bf Y}$ in the other sheet can
then be obtained by
integrating along some appropriate path in the $\rho$-$\zeta$ plane leading back
to the point $\rho_0$, $\zeta_0$, but now with $\lambda\to -\lambda$.} 
${\bf Y}$ can be discussed in general as a unique function of $\rho$, $\zeta$
and $\lambda$. Of course, each column of ${\bf Y}$ is itself a solution to the
LP.  

It can easily be verified that the following relations hold, cf.~\cite{nk83}:
\begin{equation}\label{det}
[f(\rho,\zeta)]^{-1}{\rm det}\,{\bf Y}(\rho,\zeta,\lambda)=c_0(K)\, ,
\end{equation}
\begin{equation}\label{-lambda}
{\bf Y}(\rho,\zeta,-\lambda)=\left(\begin{array}{crc}
1 & 0 & 0 \\
0 & -1 & 0 \\
0 & 0 & 1
\end{array}
\right){\bf Y}(\rho,\zeta,\lambda){\bf c}_1(K)\, ,
\end{equation}

\begin{equation}\label{CC}
\hspace{-1cm}
\left[{\bf Y}(\rho,\zeta,1/{\bar\lambda})\right]^{\dagger}
\left(\begin{array}{ccr}
[f(\rho,\zeta)]^{-1} & 0 & 0 \\
0 & -[f(\rho,\zeta)]^{-1} & 0 \\
0 & 0 & -1
\end{array}
\right){\bf Y}(\rho,\zeta,\lambda) = {\bf c}_2(K)\, ,
\end{equation}
where $c_0(K)$ as well as the $3\times 3$ matrices ${\bf c}_1(K)$ and 
${\bf c}_2(K)$ do not depend on $\rho$ and $\zeta$.

\section{Integration along axis and horizon}
\subsection{Killing horizon and axis of symmetry}\label{ha}
The event horizon of a stationary and axisymmetric black hole is given by a 
null hypersurface whose normal vector $\chi^i$ is a linear combination of the
two Killing vectors $\xi^i$ and $\eta^i$, see (\ref{killing}),
\begin{equation}\label{H1}
\chi^i\equiv \xi^i+\Omega\eta^i\, , \quad 
\mathcal H: \quad \chi^i\chi_i=0\, ,
\end{equation} 
where $\Omega$ is the constant ``angular velocity of the horizon'' 
with respect to infinity 
\cite{he73, carter73}.\footnote{The Killing vector $\xi^i$ is normalized 
by $\xi^i\xi_i\to -1$ at spatial infinity. Note that $f\equiv -\xi^i\xi_i$.}
Throughout this paper we assume
$\Omega\neq 0$. For symmetry reasons, each of the Killing vectors 
$\xi^i$ and $\eta^i$ must be tangential to the horizon, and therefore
\begin{equation}\label{H2}
\mathcal H: \quad \chi^i\xi_i=0\, , \quad \chi^i\eta_i=0\,.
\end{equation}
We assume that Weyl coordinates $\rho$ and $\zeta$, see (\ref{line}) or (\ref{line2}), 
can globally be used everywhere outside (and including) the horizon.
Because of   
\begin{equation}
\rho^2=(\xi^i\eta_i)^2-\xi^i\xi_i\eta^k\eta_k=(\chi^i\eta_i)^2-
\chi^i\chi_i\eta^k\eta_k
\end{equation}
the horizon must be located on the
$\zeta$-axis,
\begin{equation}
\mathcal H: \quad \rho=0\, .
\end{equation} 
For a single black hole, i.e.\ a connected horizon, only two possibilities
remain: 
The horizon can either be a finite interval or a point on the $\zeta$-axis, see
Figure 1. 
\begin{figure}
\unitlength1cm
\begin{picture}(8,7)
\linethickness{1mm}
\put(3.5,1.55){\line(0,1){2.9}}
\thinlines
\put(3.5,3){\vector(1,0){3}}
\put(3.5,0){\vector(0,1){6}}
\put(3.4,6.3){$\zeta$}
\put(6.7,2.9){$\rho$}
\multiput(3.6,0.45)(0,0.1){11}{\circle*{0.02}}
\multiput(3.6,5.55)(0,-0.1){11}{\circle*{0.02}}
\multiput(3.65,1.55)(0,0.1){30}{\circle*{0.02}}
\put(3.70011,5.54803){\circle*{0.02}}
\put(3.80007,5.54214){\circle*{0.02}}
\put(3.89972,5.53232){\circle*{0.02}}
\put(3.99891,5.51861){\circle*{0.02}}
\put(4.09748,5.501){\circle*{0.02}}
\put(4.19529,5.47954){\circle*{0.02}}
\put(4.29217,5.45426){\circle*{0.02}}
\put(4.38799,5.42519){\circle*{0.02}}
\put(4.4826,5.39239){\circle*{0.02}}
\put(4.57584,5.35589){\circle*{0.02}}
\put(4.66758,5.31577){\circle*{0.02}}
\put(4.75768,5.27207){\circle*{0.02}}
\put(4.84598,5.22486){\circle*{0.02}}
\put(4.93237,5.17423){\circle*{0.02}}
\put(5.0167,5.12025){\circle*{0.02}}
\put(5.09885,5.06299){\circle*{0.02}}
\put(5.17869,5.00256){\circle*{0.02}}
\put(5.25609,4.93904){\circle*{0.02}}
\put(5.33094,4.87252){\circle*{0.02}}
\put(5.40312,4.80312){\circle*{0.02}}
\put(5.47252,4.73094){\circle*{0.02}}
\put(5.53904,4.65609){\circle*{0.02}}
\put(5.60256,4.57869){\circle*{0.02}}
\put(5.66299,4.49885){\circle*{0.02}}
\put(5.72025,4.4167){\circle*{0.02}}
\put(5.77423,4.33237){\circle*{0.02}}
\put(5.82486,4.24598){\circle*{0.02}}
\put(5.87207,4.15768){\circle*{0.02}}
\put(5.91577,4.06758){\circle*{0.02}}
\put(5.95589,3.97584){\circle*{0.02}}
\put(5.99239,3.8826){\circle*{0.02}}
\put(6.02519,3.78799){\circle*{0.02}}
\put(6.05426,3.69217){\circle*{0.02}}
\put(6.07954,3.59529){\circle*{0.02}}
\put(6.101,3.49748){\circle*{0.02}}
\put(6.11861,3.39891){\circle*{0.02}}
\put(6.13232,3.29972){\circle*{0.02}}
\put(6.14214,3.20007){\circle*{0.02}}
\put(6.14803,3.10011){\circle*{0.02}}
\put(6.15,3.){\circle*{0.02}}
\put(6.14803,2.89989){\circle*{0.02}}
\put(6.14214,2.79993){\circle*{0.02}}
\put(6.13232,2.70028){\circle*{0.02}}
\put(6.11861,2.60109){\circle*{0.02}}
\put(6.101,2.50252){\circle*{0.02}}
\put(6.07954,2.40471){\circle*{0.02}}
\put(6.05426,2.30783){\circle*{0.02}}
\put(6.02519,2.21201){\circle*{0.02}}
\put(5.99239,2.1174){\circle*{0.02}}
\put(5.95589,2.02416){\circle*{0.02}}
\put(5.91577,1.93242){\circle*{0.02}}
\put(5.87207,1.84232){\circle*{0.02}}
\put(5.82486,1.75402){\circle*{0.02}}
\put(5.77423,1.66763){\circle*{0.02}}
\put(5.72025,1.5833){\circle*{0.02}}
\put(5.66299,1.50115){\circle*{0.02}}
\put(5.60256,1.42131){\circle*{0.02}}
\put(5.53904,1.34391){\circle*{0.02}}
\put(5.47252,1.26906){\circle*{0.02}}
\put(5.40312,1.19688){\circle*{0.02}}
\put(5.33094,1.12748){\circle*{0.02}}
\put(5.25609,1.06096){\circle*{0.02}}
\put(5.17869,0.997442){\circle*{0.02}}
\put(5.09885,0.937007){\circle*{0.02}}
\put(5.0167,0.879752){\circle*{0.02}}
\put(4.93237,0.825768){\circle*{0.02}}
\put(4.84598,0.775135){\circle*{0.02}}
\put(4.75768,0.727933){\circle*{0.02}}
\put(4.66758,0.684235){\circle*{0.02}}
\put(4.57584,0.644107){\circle*{0.02}}
\put(4.4826,0.607612){\circle*{0.02}}
\put(4.38799,0.574806){\circle*{0.02}}
\put(4.29217,0.545739){\circle*{0.02}}
\put(4.19529,0.520457){\circle*{0.02}}
\put(4.09748,0.498998){\circle*{0.02}}
\put(3.99891,0.481395){\circle*{0.02}}
\put(3.89972,0.467675){\circle*{0.02}}
\put(3.80007,0.457861){\circle*{0.02}}
\put(3.70011,0.451966){\circle*{0.02}}
\put(3.7,4.9){$\mathcal{A^+}$}
\put(3.7,0.9){$\mathcal{A^-}$}
\put(5.5,4.9){$\mathcal{C}$}
\put(3.1,4.35){$l$}
\put(2.75,1.4){$-l$}
\put(3.8,3.5){${\mathcal{H}}$}
\put(0.9,3.5){${\mathcal{H}}$: $\rho=0$,}
\put(1.5,2.8){$|\zeta|\le l$}
\put(8.5,3){\vector(1,0){3}}
\put(8.5,0){\vector(0,1){6}}
\put(8.4,6.3){$\zeta$}
\put(11.7,2.9){$\rho$}
\put(8.5,3){\circle*{0.3}}
\put(8.69284,3.22981){\circle*{0.02}}
\put(8.75981,3.15){\circle*{0.02}}
\put(8.79544,3.05209){\circle*{0.02}}
\put(8.79544,2.94791){\circle*{0.02}}
\put(8.75981,2.85){\circle*{0.02}}
\put(8.69284,2.77019){\circle*{0.02}}
\multiput(8.6,0.45)(0,0.099){24}{\circle*{0.02}}
\multiput(8.6,5.55)(0,-0.099){24}{\circle*{0.02}}
\put(8.70011,5.54803){\circle*{0.02}}
\put(8.80007,5.54214){\circle*{0.02}}
\put(8.89972,5.53232){\circle*{0.02}}
\put(8.99891,5.51861){\circle*{0.02}}
\put(9.09748,5.501){\circle*{0.02}}
\put(9.19529,5.47954){\circle*{0.02}}
\put(9.29217,5.45426){\circle*{0.02}}
\put(9.38799,5.42519){\circle*{0.02}}
\put(9.4826,5.39239){\circle*{0.02}}
\put(9.57584,5.35589){\circle*{0.02}}
\put(9.66758,5.31577){\circle*{0.02}}
\put(9.75768,5.27207){\circle*{0.02}}
\put(9.84598,5.22486){\circle*{0.02}}
\put(9.93237,5.17423){\circle*{0.02}}
\put(10.0167,5.12025){\circle*{0.02}}
\put(10.09885,5.06299){\circle*{0.02}}
\put(10.17869,5.00256){\circle*{0.02}}
\put(10.25609,4.93904){\circle*{0.02}}
\put(10.33094,4.87252){\circle*{0.02}}
\put(10.40312,4.80312){\circle*{0.02}}
\put(10.47252,4.73094){\circle*{0.02}}
\put(10.53904,4.65609){\circle*{0.02}}
\put(10.60256,4.57869){\circle*{0.02}}
\put(10.66299,4.49885){\circle*{0.02}}
\put(10.72025,4.4167){\circle*{0.02}}
\put(10.77423,4.33237){\circle*{0.02}}
\put(10.82486,4.24598){\circle*{0.02}}
\put(10.87207,4.15768){\circle*{0.02}}
\put(10.91577,4.06758){\circle*{0.02}}
\put(10.95589,3.97584){\circle*{0.02}}
\put(10.99239,3.8826){\circle*{0.02}}
\put(11.02519,3.78799){\circle*{0.02}}
\put(11.05426,3.69217){\circle*{0.02}}
\put(11.07954,3.59529){\circle*{0.02}}
\put(11.101,3.49748){\circle*{0.02}}
\put(11.11861,3.39891){\circle*{0.02}}
\put(11.13232,3.29972){\circle*{0.02}}
\put(11.14214,3.20007){\circle*{0.02}}
\put(11.14803,3.10011){\circle*{0.02}}
\put(11.15,3.){\circle*{0.02}}
\put(11.14803,2.89989){\circle*{0.02}}
\put(11.14214,2.79993){\circle*{0.02}}
\put(11.13232,2.70028){\circle*{0.02}}
\put(11.11861,2.60109){\circle*{0.02}}
\put(11.101,2.50252){\circle*{0.02}}
\put(11.07954,2.40471){\circle*{0.02}}
\put(11.05426,2.30783){\circle*{0.02}}
\put(11.02519,2.21201){\circle*{0.02}}
\put(10.99239,2.1174){\circle*{0.02}}
\put(10.95589,2.02416){\circle*{0.02}}
\put(10.91577,1.93242){\circle*{0.02}}
\put(10.87207,1.84232){\circle*{0.02}}
\put(10.82486,1.75402){\circle*{0.02}}
\put(10.77423,1.66763){\circle*{0.02}}
\put(10.72025,1.5833){\circle*{0.02}}
\put(10.66299,1.50115){\circle*{0.02}}
\put(10.60256,1.42131){\circle*{0.02}}
\put(10.53904,1.34391){\circle*{0.02}}
\put(10.47252,1.26906){\circle*{0.02}}
\put(10.40312,1.19688){\circle*{0.02}}
\put(10.33094,1.12748){\circle*{0.02}}
\put(10.25609,1.06096){\circle*{0.02}}
\put(10.17869,0.997442){\circle*{0.02}}
\put(10.09885,0.937007){\circle*{0.02}}
\put(10.0167,0.879752){\circle*{0.02}}
\put(9.93237,0.825768){\circle*{0.02}}
\put(9.84598,0.775135){\circle*{0.02}}
\put(9.75768,0.727933){\circle*{0.02}}
\put(9.66758,0.684235){\circle*{0.02}}
\put(9.57584,0.644107){\circle*{0.02}}
\put(9.4826,0.607612){\circle*{0.02}}
\put(9.38799,0.574806){\circle*{0.02}}
\put(9.29217,0.545739){\circle*{0.02}}
\put(9.19529,0.520457){\circle*{0.02}}
\put(9.09748,0.498998){\circle*{0.02}}
\put(8.99891,0.481395){\circle*{0.02}}
\put(8.89972,0.467675){\circle*{0.02}}
\put(8.80007,0.457861){\circle*{0.02}}
\put(8.70011,0.451966){\circle*{0.02}}
\put(8.8,3.3){$\mathcal{H}$}
\put(8.7,4.9){$\mathcal{A^+}$}
\put(8.7,0.9){$\mathcal{A^-}$}
\put(10.5,4.9){$\mathcal{C}$}
\put(12,5.2){$\rho=r\sin\theta$}
\put(12,4.5){$\zeta=r\cos\theta$}
\put(12,3.8){$(0\le\theta\le\pi)$}
\put(12,2.0){$\mathcal{H}$: $r=0,$}
\put(12.75,1.3) {$0\le\theta\le\pi$}
\end{picture}
\caption{The two possibilities for a connected horizon in Weyl coordinates: 
a finite interval on the $\zeta$-axis (left picture) and a point on the
$\zeta$-axis (right picture).}
\label{fig}
\end{figure}
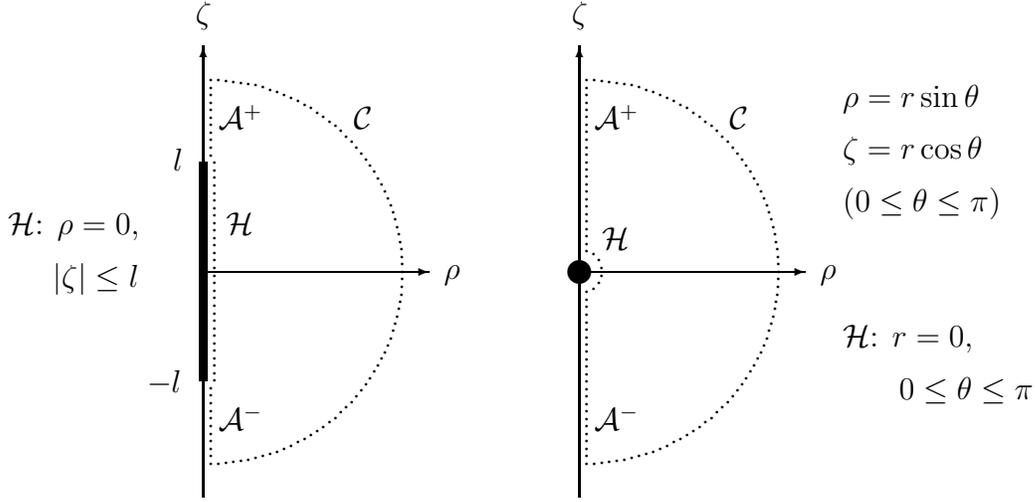
Without loss of generality, we have placed the horizon in a symmetrical
position 
with respect to $\zeta=0$. In the case of a point on the $\zeta$-axis (here
$\zeta=0$) the slice $t={\rm constant}$, $\phi={\rm constant}$ of the horizon
still has to be one-dimensional, of course. Therefore, we have to parametrize
the position along the horizon by another coordinate. A possible choice is the
angle $\theta$ of spherical-like coordinates (\ref{sc}).

On the axis of symmetry, $\rho=0$ holds because the Killing vector 
$\boldsymbol\eta$ vanishes there. We denote the two parts of the axis of
symmetry by $\mathcal{A^+}$ and $\mathcal{A^-}$, see Figure 1. The metric
function $a$, see (\ref{line}), vanishes on $\mathcal{A^{\pm}}$:
\begin{equation}\label{aA}
\mathcal{A^{\pm}}: \quad a=0\, . 
\end{equation}
On $\mathcal{H}$, however, because of  
\begin{equation}
\chi^i\chi_i \equiv -f[(1+\Omega a)^2-\Omega^2\rho^2f^{-2}]=0\, ,
\end{equation}
we have
\begin{equation}\label{aH}
\mathcal{H}: \quad a=-\frac{1}{\Omega}\, . 
\end{equation}
Note that $f\neq 0$ holds on the horizon of rotating black holes except at the
``poles'' (the points where the horizon meets the axis of symmetry), since 
$f=-\Omega^2\eta^i\eta_i$ follows from (\ref{H1}, \ref{H2}), and for the
spacelike vector $\boldsymbol\eta$ we have $\eta^i\eta_i>0$ everywhere except on
the axis of symmetry. We emphasize that despite the discontinuity of $a$ at the
poles of the horizon all scalar products $\eta^i\eta_i$, $\xi^i\xi_i$ and
$\xi^i\eta_i$ and thus the metric coefficients $g_{\phi\phi}$, $g_{tt}$ and
$g_{\phi t}$ are, of course, continuous there. We assume that the complex
potentials $\mathcal E$ and $\Phi$ as well as the solution ${\bf Y}$ to the LP
are continuous as well. 

\subsection{Integration of the Linear Problem for $\rho=0$}
For $\rho=0$, i.e.\ on $\mathcal{A^+}$, $\mathcal{A^-}$ and  $\mathcal H$, 
the function $\lambda(K)$ as given in (\ref{lambda}) degenerates: The two branch
points $K={\rm i}\bar z$ and $K=-{\rm i}z$ merge to $K=\zeta$. As long as 
$K\neq\zeta$, we simply get $\lambda=\pm 1$. 
Along any curve where $\rho=0$ holds, the LP (\ref{LP1}, \ref{LP2}) can
easily 
be integrated. The general solution for $\lambda=+1$ is given by\footnote{The
structure of the three columns of the first matrix on the right-hand side of
(\ref{+1}) can also be found in \cite{nk83} when the transformation
(\ref{trans}) is taken into account.} 
\begin{equation}\label{+1}
{\bf Y}^{(+1)}(\tau,K)=\left(\begin{array}{crr}
\bar{\mathcal E}(\tau)+2|\Phi(\tau)|^2 & 1 & \Phi(\tau) \\
{\mathcal E(\tau)} & -1 & -\Phi(\tau) \\
2\bar \Phi(\tau) & 0 & 1
\end{array}
\right){\bf C}(K)\, ,
\end{equation}
where $\tau$ stands here for $\zeta$ on $\mathcal{A^+}$, $\mathcal{A^-}$ or 
$\mathcal H$, if the horizon is given by a finite interval on the $\zeta$-axis.
For a horizon placed at the origin of the $\rho$-$\zeta$ plane we may identify
$\tau$ with $\theta$. Note that the LP (\ref{LP1}, \ref{LP2}) can be
reformulated as a pair of equations for 
${\bf Y}_{,\rho}$ and ${\bf Y}_{,\zeta}$ as well as for ${\bf Y}_{,r}$ and 
${\bf Y}_{,\theta}$. We denote the solutions ${\bf Y}^{(+1)}(\tau,K)$ along 
$\mathcal{A^{\pm}}$ by ${\bf Y}_{\pm}$ and along 
$\mathcal H$ by ${\bf Y}_{\rm h}$ as well as the corresponding 
matrices ${\bf C}(K)$ by ${\bf C}_{\pm}$ and ${\bf C}_{\rm h}$:
\begin{equation}\label{Y+-}
\mathcal{A^{\pm}}: \quad {\bf Y}^{(+1)}(\zeta,K)\equiv {\bf Y}_{\pm}=
\left(\begin{array}{crr}
\bar{\mathcal E}+2|\Phi|^2 & 1 & \Phi \\
{\mathcal E} &-1 & -\Phi \\
2\bar \Phi & 0 & 1
\end{array}
\right){\bf C}_{\pm}\, ,
\end{equation}
\begin{equation}\label{Yh}
\mathcal H: \hspace{0.7cm}  {\bf Y}^{(+1)}(\tau,K)\equiv {\bf Y}_{\rm h}=
\left(\begin{array}{crr}
\bar{\mathcal E}+2|\Phi|^2 & 1 & \Phi \\
{\mathcal E} &-1 & -\Phi \\
2\bar \Phi & 0 & 1
\end{array}
\right){\bf C}_{\rm h}\, .
\end{equation}  
As already discussed in Section 2, a particular solution to the LP for given
$\mathcal E$, $\Phi$ is fixed by prescribing ${\bf Y}$ as a function of $K$ at
some point $\rho=\rho_0$, $\zeta=\zeta_0$ in one of the two sheets of the
Riemann $K$-surface. This represents the ``initial condition'' for integrating
the LP along any path in the $\rho$-$\zeta$ plane. Choosing some point
$\zeta=\zeta_0$ on $\mathcal{A^+}$ as the ``starting point'' and taking
$K\neq\zeta$, $\lambda=+1$ there, this corresponds directly to prescribing the
matrix
function ${\bf C}_+(K)$ in (\ref{Y+-}). (In general such an initial condition,
which can 
also be called ``normalization'', removes the freedom of multiplying ${\bf Y}$
from the right with a matrix not depending on the coordinates.) The matrices
${\bf C}_-(K)$ and ${\bf C}_{\rm h}(K)$
are then determined by continuity conditions, see
below. We define the matrix ${\bf C}_+(K)$ by setting
\begin{equation}\label{defC+}
\lim_{K\to \zeta}{\bf Y}_+(\zeta,K) =\left(\begin{array}{crc}
1 & 1 & 0 \\
1 & -1 & 0 \\
0 & 0 & 1                                        
\end{array}\right)
\end{equation}
together with the requirement that the elements of ${\bf C}_+(K)$ be
holomorphic functions of $K$. With (\ref{Y+-}) this leads to
\begin{equation}\label{C+}
{\bf C}_+ = \left(\begin{array}{ccc}
F & 0 & 0 \\
G & 1 & L \\
H & 0 & 1
\end{array}\right)\, ,
\end{equation}
where the functions $F(K)$, $G(K)$, $H(K)$ and $L(K)$, for $K=\zeta$, are given 
by the potentials $\mathcal E=\mathcal E_+$, $\Phi=\Phi_+$ on $\mathcal{A^+}$:
\begin{equation}\label{F}
F(\zeta)=[f_+(\zeta)]^{-1}\, ,
\end{equation}
\begin{equation}
G(\zeta)=\left[|\Phi_+(\zeta)|^2+{\rm i}b_+(\zeta)\right][f_+(\zeta)]^{-1}\, ,
\end{equation}
\begin{equation}
H(\zeta)=-2\bar\Phi_+(\zeta)[f_+(\zeta)]^{-1}\, ,
\end{equation}
\begin{equation}\label{L}
L(\zeta)=-\Phi_+(\zeta)\, .
\end{equation}
Here we have used the function $f$ as defined in (\ref{f}) and the abbreviation
\begin{equation}
b\equiv \Im\,\mathcal E
\end{equation}
for the imaginary part of $\mathcal E$. Vice versa, the axis potentials can be
expressed as
\begin{equation}\label{ax}
\mathcal E_+(\zeta)=\frac{1-\bar G(\zeta)}{F(\zeta)}\, , \quad
\Phi_+(\zeta)=-\frac{\bar H(\zeta)}{2F(\zeta)}\,.
\end{equation}
The functions $F$, $G$, $H$, $L$ for all (complex) $K$ are given by the
analytic 
continuation of the expressions (\ref{F}-\ref{L}) in terms of the axis
potentials. These results are generalizations of corresponding formulae
published in \cite{mn95}. The following important properties can immediately be
seen:
\begin{equation}\label{cc}
\hspace{-1cm}\overline{F(\bar K)}=F(K)\, , \quad \overline{G(\bar K)}+G(K)=
\frac{\overline{H(\bar K)}H(K)}{2F(K)}\, , \quad 
L(K)=\frac{\overline{H(\bar K)}}{2F(K)}\, .
\end{equation}
The last equation shows that $L$ can be determined from $H$ and $F$, i.e.\ all 
information is already contained in the three functions $F$, $G$ and $H$, see
also (\ref{ax}).

We are now in a position to calculate $c_0(K)$, ${\bf c}_1(K)$ and 
${\bf c}_2(K)$ of (\ref{det}-\ref{CC}) for our normalization of the solution 
to the LP. From (\ref{Y+-}) and (\ref{C+}) we obtain $c_0(K)=-2F(K)$, i.e.\
\begin{equation}\label{det2}
{\rm det}\,{\bf Y}(\rho,\zeta,\lambda)=-2f(\rho,\zeta)F(K)\,.
\end{equation}
The matrix  ${\bf c}_1(K)$ in (\ref{-lambda}) can easily be obtained by choosing
one of the
 branch points $K={\rm i}\bar z$ and $K=-{\rm i}z$, i.e.\ $\lambda=0$ or 
$\lambda=\infty$, where ${\bf Y}$ is a unique function of $K$, 
${\bf Y}(\rho,\zeta,-\lambda)={\bf Y}(\rho,\zeta,\lambda)$, a property 
that remains valid on  $\mathcal{A^+}$ when taking the limit $K\to\zeta$. With
(\ref{defC+}) 
we find for $K=\zeta$ 
\begin{equation*}
{\bf c}_1=\left(\begin{array}{ccc}
0 & 1 & 0 \\
1 & 0 & 0 \\
0 & 0 & 1
\end{array}\right)\, ,
\end{equation*}
which must hold for all $K$ by analytic continuation. Hence we have
\begin{equation}\label{-lambda2}
{\bf Y}(\rho,\zeta,-\lambda)=\left(\begin{array}{crc}
1 & 0 & 0 \\
0 & -1 & 0 \\
0 & 0 & 1
\end{array}
\right){\bf Y}(\rho,\zeta,\lambda)\left(\begin{array}{ccc}
0 & 1 & 0 \\
1 & 0 & 0 \\
0 & 0 & 1
\end{array}\right)\, .
\end{equation} 
Finally, ${\bf c}_2(K)$ in (\ref{CC}) can be calculated in a straightforward
manner\footnote{It is
sufficient to consider (\ref{Y+-}) in the limit $\zeta\to\infty$ making use of
$\mathcal E\to 1$ and $\Phi\to 0$, see (\ref{as1}, \ref{as2}).} 
from (\ref{Y+-}) and (\ref{C+}) together with (\ref{cc}) leading to
\begin{equation}
\hspace{-1cm} \left[{\bf Y}(\rho,\zeta,1/{\bar\lambda})\right]^{\dagger}
\left(\begin{array}{ccr}
f^{-1} & 0 & 0 \\
0 & -f^{-1} & 0 \\
0 & 0 & -1
\end{array}
\right){\bf Y}(\rho,\zeta,\lambda) = \left(\begin{array}{ccr}
0 & 2F & 0 \\
2F & 0 & 0 \\
0 & 0 & -1
\end{array}\right)\, .
\end{equation}
Note that (\ref{cc}) can also be written in matrix form:
\begin{equation}\label{CCK}
\left[{\bf C}_+(\bar K)\right]^{\dagger}\left(\begin{array}{ccr}
0 & 2 & 0 \\
2 & 0 & 0 \\
0 & 0 & -1                                    
\end{array}\right){\bf C}_+(K)=\left(\begin{array}{ccr}
0 & 2F & 0 \\
2F & 0 & 0 \\
0 & 0 & -1
\end{array}\right)\, .
\end{equation}

\subsection{The corotating frame of reference}
As already mentioned above, the matrices ${\bf C}_-(K)$ and ${\bf C}_{\rm h}(K)$
follow from the prescribed ${\bf C}_+(K)$ by continuity conditions at the poles
of the horizon, i.e.\ at $\zeta=\pm l$ or $\theta=0,\pi$; see Figure 1.
However, as discussed at the end of Subsection 3.1, the function $f=-\xi^i\xi_i$
vanishes there:
\begin{equation}\label{fns}
f_{\rm n}=0\, , f_{\rm s}=0\, . 
\end{equation}
(From here on we use the index ``n'' to denote quantities at the ``north pole'',
where $\mathcal H$ and $\mathcal{A^+}$ meet, and ``s'' for quantities at the
``south pole'', where $\mathcal H$ and $\mathcal{A^-}$ meet.) Because of
(\ref{det2}), this means that ${\rm det}\,{\bf Y}$ vanishes at the poles.
Therefore, the continuity of ${\bf Y}_+$ does not provide enough information.
This can be compensated by considering the continuity of 
${\bf Y}'_+$ in addition, where ${\bf Y}'$
denotes a solution to the LP in the corotating frame of reference defined by
\begin{equation}
\rho'=\rho\, , \quad \zeta'=\zeta\, , \quad \phi'=\phi-\Omega t\, , \quad
t'=t\, .
\end{equation}
It has been shown in \cite{ha09}, by generalizing the corresponding
transformation in the pure vacuum case (i.e.\ without electromagnetic field),
see \cite{nm03}, that
${\bf Y}'$ can be expressed as\footnote{When comparing our expression with the
formula in \cite{ha09} one has to be aware of (\ref{trans}). Note that the
third line of ${\bf Y}'$ is identical to that of ${\bf Y}$.}
\begin{equation}
{\bf Y}'(\rho,\zeta,\lambda)=
{\bf T}(\rho,\zeta,\lambda){\bf Y}(\rho,\zeta,\lambda)\, , 
\end{equation}
\begin{equation}
{\bf T}=\left(\begin{array}{ccc}
x & 0 & 0 \\
0 & y & 0 \\
0 & 0 & 1
\end{array}\right) + {\rm i}(K+{\rm i}z)\Omega f^{-1}\left(
\begin{array}{rrc}
-1 & -\lambda & 0 \\
\lambda & 1 & 0 \\
0 & 0 & 0 
\end{array}\right)
\end{equation}
with
\begin{equation}
x=1+\Omega(a-\rho f^{-1})\, , \quad y=1+\Omega(a+\rho f^{-1})\, .
\end{equation}
Together with (\ref{aA}) and (\ref{aH}) this leads to the following formulae
for ${\bf Y}_{\pm}'$ and  ${\bf Y}_{\rm h}'$:
\begin{equation}
\hspace{-1.5cm}
{\bf Y}_{\pm}'(\zeta,K)={\bf Y}_{\pm}(\zeta,K)+2{\rm i}(K-\zeta)\Omega\left(
\begin{array}{rcc}
-1 & 0 & 0 \\
1 & 0 & 0 \\
0 & 0 & 0 
\end{array}\right){\bf C}_{\pm}(K)\, ;
\end{equation}
\begin{equation}
\hspace{-1.5cm}
{\bf Y}_{\rm h}'(\zeta,K)=\left(
\begin{array}{ccc}
0 & 0 & 0 \\
0 & 0 & 0 \\
0 & 0 & 1 
\end{array}\right)
{\bf Y}_{\rm h}(\zeta,K) + 2{\rm i}(K-\zeta)\Omega\left(
\begin{array}{rcc}
-1 & 0 & 0 \\
1 & 0 & 0 \\
0 & 0 & 0 
\end{array}\right){\bf C}_{\rm h}(K)\, , 
\end{equation}
if the horizon is located at $\rho=0$, $|\zeta|\le l$, and
\begin{equation}
\hspace{-1.5cm}
{\bf Y}_{\rm h}'(\theta,K)=\left(
\begin{array}{ccc}
0 & 0 & 0 \\
0 & 0 & 0 \\
0 & 0 & 1 
\end{array}\right)
{\bf Y}_{\rm h}(\theta,K) + 2{\rm i}K\Omega\left(
\begin{array}{rcc}
-1 & 0 & 0 \\
1 & 0 & 0 \\
0 & 0 & 0 
\end{array}\right){\bf C}_{\rm h}(K)\, , 
\end{equation}
if the horizon is located at $r=0$. Combining these
formulae with (\ref{Y+-}, \ref{Yh}) and making use of (\ref{fns}) we get the
following continuity conditions at the poles that permit the calculation of
${\bf C}_-(K)$ and ${\bf C}_{\rm h}(K)$ for a given ${\bf C}_+(K)$:
\begin{equation}\label{nn}
{\bf A}_{\rm n}{\bf C}_+={\bf H}_{\rm n}{\bf C}_{\rm h}\, , 
\end{equation}
\begin{equation}\label{ss}
{\bf A}_{\rm s}{\bf C}_-={\bf H}_{\rm s}{\bf C}_{\rm h}\, , 
\end{equation}
where the matrices ${\bf A}_{\rm n}$, ${\bf H}_{\rm n}$, 
${\bf A}_{\rm s}$  and ${\bf H}_{\rm s}$ depend on the values of
$\mathcal E$ and $\Phi$ at the poles:
\begin{equation}
{\bf A}_{\rm n/s}=\left(\begin{array}{crc}
\mathcal E_{\rm n/s} & -1 & -\Phi_{\rm n/s} \\
\mathcal E_{\rm n/s}+2{\rm i}(K\mp l)\Omega & -1 & -\Phi_{\rm n/s} \\
2\bar \Phi_{\rm n/s} & 0 & 1
\end{array}\right)\, ,
\end{equation}
\begin{equation}
{\bf H}_{\rm n/s}=\left(\begin{array}{crc}
\mathcal E_{\rm n/s} & -1 & -\Phi_{\rm n/s} \\
2{\rm i}(K\mp l)\Omega & 0 & 0 \\
2\bar \Phi_{\rm n/s} & 0 & 1
\end{array}\right)\, . 
\end{equation}
These formulae are also valid when the horizon is located at $r=0$. In this
case one simply has to put $l=0$.

\section{Proof of uniqueness}
For $\zeta\to \pm\infty$, we obtain from (\ref{Y+-}) using $\mathcal E\to 1$ and
$\Phi\to 0$, see 
(\ref{as1}, \ref{as2}),
\begin{equation}
{\bf Y}_{\pm}\to\left(\begin{array}{crc}
1 & 1 & 0 \\
1 & -1 & 0 \\
0 & 0 & 1 
\end{array}\right){\bf C}_{\pm}\, .
\end{equation}
Closing the integration path via infinity, say along a half circle
$\rho=R\sin\theta$, $\zeta=R\cos\theta$ ($0\le\theta\le\pi$, $R\to\infty$) as
indicated by the curve $\mathcal C$ in Figure 1, we can conclude that
\begin{equation}
\hspace{-1.5cm}
\left(\begin{array}{crc}
1 & 1 & 0 \\
1 & -1 & 0 \\
0 & 0 & 1 
\end{array}\right){\bf C}_- =
\left(\begin{array}{crc}
1 & 0 & 0 \\
0 & -1 & 0 \\
0 & 0 & 1
\end{array}
\right)\left(\begin{array}{crc}
1 & 1 & 0 \\
1 & -1 & 0 \\
0 & 0 & 1 
\end{array}\right){\bf C}_+
\left(\begin{array}{ccc}
0 & 1 & 0 \\
1 & 0 & 0 \\
0 & 0 & 1
\end{array}\right)\, . 
\end{equation}
This follows from (\ref{-lambda2}) and the fact that ${\bf Y}$ does not change
along $\mathcal C$ since all coefficients $A_i$, $B_i$, $C_i$, $D_i$ ($i=1,2$)
in (\ref{LP1}, \ref{LP2}) as defined in (\ref{AB}, \ref{CD}) vanish
sufficiently rapidly as $R\to\infty$ because of (\ref{as1}, \ref{as2}), but
$\lambda$, according to (\ref{lambda}), changes from $\pm 1$ at $\theta=0$ to
$\mp 1$ at $\theta=\pi$
\cite{mn95}. Thus
\begin{equation}\label{C+-}
{\bf C}_- =  \left(\begin{array}{ccc}
0 & 1 & 0 \\
1 & 0 & 0 \\
0 & 0 & 1
\end{array}\right){\bf C}_+
\left(\begin{array}{ccc}
0 & 1 & 0 \\
1 & 0 & 0 \\
0 & 0 & 1
\end{array}\right)\, . 
\end{equation}
Together with (\ref{nn}, \ref{ss}) this relation already permits the explicit
calculation of the elements of ${\bf C}_+$ as functions of $K$ and the
parameters $\Omega$, $l$, $\mathcal E_{\rm n}$, $\mathcal E_{\rm s}$, 
$\Phi_{\rm n}$ and $\Phi_{\rm s}$. To this end, we define the matrix
\begin{equation}\label{M}
{\bf M} \equiv {\bf C}_+({\bf C}_-)^{-1}{\bf m}_0 \quad \mbox{with} \quad 
{\bf m}_0=
\left(\begin{array}{ccc}
1 & 0 & 0 \\
0 & 1 & 0 \\
0 & 0 & -2
\end{array}\right)\, , 
\end{equation}
which, according to (\ref{C+}), can be written in terms of the functions $F(K)$,
$G(K)$, $H(K)$ and $L(K)$ as
\begin{equation}\label{M2}
{\bf M} = \left(\begin{array}{ccc}
F & HL-G & 2FL \\
G & (1-G)(1+G-HL)/F & 2L(G-1) \\
H & H(HL-G-1)/F & 2(HL-1)
\end{array}\right)\, .
\end{equation}
From (\ref{CCK}), (\ref{C+-}) and (\ref{M}) the remarkable property
\begin{equation}\label{Mcc}
[{\bf M}(\bar K)]^{\dagger} = {\bf M}(K)
\end{equation}
follows, which can also be checked directly in (\ref{M2}) by means of
(\ref{cc}). According to its definition (\ref{M}) and (\ref{nn}, \ref{ss}), the
matrix ${\bf M}$ can be calculated as
\begin{equation}
{\bf M} = {\bf A}_{\rm n}^{-1}{\bf H}_{\rm n}{\bf H}_{\rm s}^{-1} 
{\bf A}_{\rm s}{\bf m}_0 \, ,
\end{equation}
leading to
\begin{equation}
{\bf M} = \left({\bf 1} + \frac{{\bf F}_{\rm n}}{2{\rm i}\Omega(K-l)}\right)
\left({\bf 1} - \frac{{\bf F}_{\rm s}}{2{\rm i}\Omega(K+l)}\right){\bf m}_0
\end{equation}
with
\begin{equation}
{\bf F}_{\rm n/s} = \left(\begin{array}{ccc}
-{\mathcal E}_{\rm n/s} & 1 & \Phi_{\rm n/s} \\
|{\mathcal E}_{\rm n/s}|^2 & -\bar{\mathcal E}_{\rm n/s} & 
-\Phi_{\rm n/s}\bar{\mathcal E}_{\rm n/s} \\
2\bar\Phi_{\rm n/s}{\mathcal E}_{\rm n/s}  & 
-2\bar\Phi_{\rm n/s} & -2|\Phi_{\rm n/s}|^2
\end{array}\right)\, .
\end{equation}
Obviously, the matrix {\bf M} can also be written in the form
\begin{equation}
{\bf M}={\bf m}_0+\frac{{\bf m}_1}{2{\rm i}\Omega(K-l)}+
\frac{{\bf m}_2}{2{\rm i}\Omega(K+l)}+
\frac{{\bf m}_3}{4\Omega^2(K^2-l^2)}
\end{equation}
with  
\begin{equation}
{\bf m}_1={\bf F}_{\rm n}{\bf m}_0 \, , \quad
{\bf m}_2=-{\bf F}_{\rm s}{\bf m}_0 \, , \quad
{\bf m}_3={\bf F}_{\rm n}{\bf F}_{\rm s}{\bf m}_0 \, .
\end{equation}
The constant matrices ${\bf m}_i$ ($i=1,2,3$) are solely determined by the
parameters $\mathcal E_{\rm n}$, $\mathcal E_{\rm s}$, 
$\Phi_{\rm n}$ and $\Phi_{\rm s}$. Together with $\Omega$ and $l$ this corresponds, 
because of (\ref{fns}), to eight free real parameters. However, the property 
(\ref{Mcc}) leads to the constraints
\begin{equation}\label{cons1}
{\bf m}_1 + {\bf m}_1^{\dagger} = -({\bf m}_2 + {\bf m}_2^{\dagger})\, ,
\end{equation}
\begin{equation}\label{cons2}
4{\rm i}\Omega l({\bf m}_1 + {\bf m}_1^{\dagger})=
{\bf m}_3 - {\bf m}_3^{\dagger} \, ,
\end{equation}
which reduce the number of free real parameters to four. The evaluation of
(\ref{cons1}, \ref{cons2}) results in the parameter relations
\begin{equation}
{\mathcal E}_{\rm s}=\bar {\mathcal E}_{\rm n} \, , \quad 
\Phi_{\rm s}=
\Phi_{\rm n}\,\frac{1-\bar {\mathcal E}_{\rm n}}{1-{\mathcal E}_{\rm n}} \, ,
\end{equation}
\begin{equation}
4\Omega l = \frac{{\rm i}({\mathcal E}_{\rm n}-
\bar {\mathcal E}_{\rm n})(1-|{\mathcal E}_{\rm n}|^2)}
{|1-{\mathcal E}_{\rm n}|^2}\, .
\end{equation}
Now we can read off the functions $F(K)$, $G(K)$ and $H(K)$ from the first
column of ${\bf M}$, see (\ref{M2}), and get, for $K=\zeta$, the complex Ernst
potentials on ${\mathcal A}^+$ according to (\ref{ax}). From (\ref{as1},
\ref{as2}), evaluated for $\theta=0$ (i.e.\ $r=\zeta$), we can calculate the
multipole moments $M$, $J$ and $Q$, where the condition that the coefficient of
the $\zeta^{-1}$-term in (\ref{as2}) has to be real leads to the additional
parameter constraint
\begin{equation}
\Re\, \Phi_{\rm s} = \Re\, \Phi_{\rm n}\, , 
\end{equation}
which ensures the absence of a magnetic monopole and reduces the number of free
real parameters to the final value of three. The potentials
$\mathcal E_+(\zeta)$ and $\Phi_+(\zeta)$ can most easily be expressed in terms
of $M$, $J$ and $Q$. The result is
\begin{equation}\label{result}
\mathcal E_+=1-\frac{2M}{\zeta+M-{\rm i}J/M} \, , \quad 
\Phi_+=\frac{Q}{\zeta+M-{\rm i}J/M} 
\end{equation}
together with the parameter relations
\begin{equation}\label{P1}
\frac{l^2}{M^2}+\frac{Q^2}{M^2}+\frac{J^2}{M^4}=1
\end{equation}
and
\begin{equation}
\Omega M=\frac{J/M^2}{(1+l/M)^2+J^2/M^4}\, .
\end{equation}
It is well-known that the axis potentials $\mathcal E_+(\zeta)$ and
$\Phi_+(\zeta)$ 
fix the solution uniquely, cf.\ \cite{he81}. 
Since (\ref{result}) represents the axis potentials of the Kerr-Newman
solution, 
the uniqueness proof is completed. The solution for all $\rho$ and $\zeta$ is
given by
\begin{equation}\label{knfull}
\mathcal E=1-\frac{2M}{\tilde r-{\rm i}(J/M)\cos \tilde\theta} \, , \quad 
\Phi=\frac{Q}{\tilde r-{\rm i}(J/M)\cos \tilde\theta} \, ,
\end{equation} 
where the Boyer-Lindquist coordinates 
$\tilde r$ and $\tilde\theta$ are related to our Weyl coordinates $\rho$ 
and $\zeta$ by
\begin{equation}\label{BL}
\rho=\sqrt{\tilde r^2-2M\tilde r+J^2/M^2+Q^2}\,\sin\tilde\theta \, , \quad
\zeta=(\tilde r-M)\cos\tilde\theta \, ,
\end{equation}
see, for example, \cite{es}. Note that we assume $M>0$, of 
course.\footnote{It can easily be seen from (\ref{result}) that the transition
$M\to -M$ corresponds to 
$\mathcal E\to\bar \mathcal E^{-1}$, $\Phi\to\bar \mathcal E^{-1}\bar\Phi$, 
which is a special case of the invariance transformations \cite{nk69} combined
with complex conjugation. From this one can conclude that the negative $M$
solutions are singular at the ring 
$\rho=\sqrt{J^2/M^2+Q^2}$, $\zeta=0$, where $\mathcal E$ of the positive 
$M$ solution vanishes. All this can also be verified directly from
(\ref{knfull}) and (\ref{BL}), of course. Note that $M=0$ is incompatible with
the parameter constraints (\ref{cons1}, \ref{cons2}) and (\ref{fns}).} 
The relation (\ref{P1}) shows that
$Q^2/M^2 + J^2/M^4<1$
holds if the horizon is located on a finite interval of the $\zeta$-axis 
($l>0$). When the horizon is a point on the $\zeta$-axis, we have to put $l=0$,
cf.\ the remark at the end of Section 3, i.e., we are uniquely led to the
extremal Kerr-Newman black hole with 
$Q^2/M^2 + J^2/M^4=1$. 

\section{Some remarks on reflection symmetry}
The Kerr-Newman solution shows reflection symmetry with respect 
to the ``equatorial plane'' $\zeta=0$:
\begin{equation}\label{symm1}
\mathcal E(\rho,-\zeta) = \bar\mathcal E(\rho,\zeta)\, , \quad
\Phi(\rho,-\zeta) = \bar\Phi(\rho,\zeta)\, .
\end{equation}
It has been shown in \cite{ps06, emr06} that this symmetry, for any 
asymptotically flat solution, is equivalent to the properties
\begin{equation}\label{symm2}
\mathcal E_+(\zeta)\bar\mathcal E_+(-\zeta)=1 \, , \quad
\Phi_+(\zeta)=-\bar\Phi_+(-\zeta)\mathcal E_+(\zeta)
\end{equation} 
of the axis potentials (satisfied on a part of the axis extending to 
$\zeta\to +\infty$). In the pure vacuum case, this reduces to the relation 
derived in \cite{mn95, kordas95}.
As a byproduct of the results of the present paper, we can give a short 
independent proof of (\ref{symm2}): Using (\ref{Y+-}), (\ref{C+}),
(\ref{-lambda2}) and (\ref{C+-}) for $K=\zeta$, we obtain the following
expressions for the axis potentials on $\mathcal A^-$:
\begin{equation}
\mathcal E_-(\zeta)=\frac{F(\zeta)}{1-\bar G(\zeta)}\, , \quad
\Phi_-(\zeta)=\frac{\bar H(\zeta)}{2[1-\bar G(\zeta)]} \,.
\end{equation} 
Together with the previously derived expressions (\ref{ax}) on 
$\mathcal A^+$, the symmetry relations (\ref{symm1}) imply
\begin{equation}\label{FGH+-}
\frac{F(\zeta)}{1-\bar G(\zeta)}=\frac{1-G(-\zeta)}{F(-\zeta)} \, , \quad
\frac{\bar H(\zeta)}{1-\bar G(\zeta)}=-\frac{H(-\zeta)}{F(-\zeta)} \, ,
\end{equation}
where we have used that $F(\zeta)$ is real. Using (\ref{ax}) again, these 
relations turn out to be identical to (\ref{symm2}). Note that the horizon did
not play any role in these conclusions.

By means of a duality rotation $\Phi\to\Phi\exp({\rm i}\delta)$ one can 
easily see that the slightly more general case of a symmetry
\begin{equation}
\mathcal E(\rho,-\zeta) = \bar\mathcal E(\rho,\zeta)\, , \quad
\Phi(\rho,-\zeta) = {\rm e}^{2{\rm i}\delta}\,\bar\Phi(\rho,\zeta)
\end{equation}
($\delta$ being a real constant) is characterized by the axis relations
\begin{equation}
\mathcal E_+(\zeta)\bar\mathcal E_+(-\zeta)=1 \, , \quad
\Phi_+(\zeta)=
-{\rm e}^{2{\rm i}\delta}\,\bar\Phi_+(-\zeta)\mathcal E_+(\zeta) \, ,
\end{equation}
see also \cite{emr06}.  

\section{Discussion}
The method to derive the complex Ernst potentials on the axis of symmetry from
the boundary conditions at the horizon used in the present paper is a
straightforward generalization of the method that was originally developed for
solving the boundary value problem to the vacuum Einstein equations for a
rigidly rotating disc of dust \cite{nm95}. In the pure vacuum case, the method
was also applied to derive the Kerr metric as the unique solution describing a
single black hole, see Section 1, and to treat the equilibrium problem of two
rotating black holes leading to the recent non-existence proof \cite{nh11}. With
the extension of the method to the Einstein-Maxwell case further interesting
applications like the investigation of rotating discs of charged dust or
equilibrium configurations of two rotating charged black holes become
treatable. 

\ack
The author would like to thank Gernot Neugebauer for many valuable discussions.

\end{document}